\documentclass[sigconf]{acmart}
\AtBeginDocument{%
  \providecommand\BibTeX{{%
    \normalfont B\kern-0.5em{\scshape i\kern-0.25em b}\kern-0.8em\TeX}}}

\copyrightyear{2024}
\acmYear{2024}
\setcopyright{rightsretained}
\acmConference[AHs 2024]{The Augmented Humans International Conference}{April 4--6, 2024}{Melbourne, VIC, Australia}
\acmBooktitle{The Augmented Humans International Conference (AHs 2024), April 4--6, 2024, Melbourne, VIC, Australia}
\acmDOI{10.1145/3652920.3653043}
\acmISBN{979-8-4007-0980-7/24/04}

\acmConference[AHs '24]{Augmented Humans International Conference}{April 4--6,
  2024}{Melbourne, AU}
\acmBooktitle{AHs '24: Augmented Humans International Conference,
 April 4--6, 2024, Melbourne, AU} 
\acmISBN{979-8-4007-0980-7/24/04}

\begin{document}

\title{Putting Our Minds Together: Iterative Exploration for Collaborative Mind Mapping}

\author{Ying Yang}
\email{ying.yang@monash.edu}
\affiliation{%
  \institution{Monash University}
  \streetaddress{Wellington Road, Clayton}
  \city{Melbourne}
  \state{Victoria}
  \country{Australia}
  \postcode{3800}
}

\author{Tim Dwyer}
\affiliation{%
  \institution{Monash University}
  \city{Melbourne}
  \country{Australia}}
\email{tim.dwyer@monash.edu}

\author{Zachari Swiecki}
\affiliation{%
  \institution{Monash University}
  \city{Melbourne}
  \country{Australia}}
\email{zach.swiecki@monash.edu}

\author{Benjamin Lee}
\affiliation{%
  \institution{University of Stuttgart}
  \city{Stuttgart}
  \country{Germany}
}
\email{benjamin.lee@visus.uni-stuttgart.de}

\author{Michael Wybrow}
\affiliation{%
  \institution{Monash University}
  \city{Melbourne}
  \country{Australia}
}
\email{michael.wybrow@monash.edu}

\author{Maxime Cordeil}
\affiliation{%
  \institution{The University of Queensland}
  \streetaddress{St Lucia}
  \city{Brisbane}
  \state{Queensland}
  \country{Australia}
  \postcode{4072}
}
\email{m.cordeil@uq.edu.au}

\author{Teresa Wulandari}
\affiliation{%
  \institution{The University of Melbourne}
  \city{Melbourne}
  \country{Australia}
}
\email{teresa.wulandari@unimelb.edu.au}

\author{Bruce H. Thomas}
\affiliation{%
  \institution{University of South Australia}
  \streetaddress{Mawson Lakes Blvd}
  \city{Adelaide}
  \state{South Australia}
  \country{Australia}
  \postcode{5095}
}
\email{bruce.thomas@unisa.edu.au}

\author{Mark Billinghurst}
\affiliation{%
  \institution{University of South Australia}
  \streetaddress{Mawson Lakes Blvd}
  \city{Adelaide}
  \state{South Australia}
  \country{Australia}
  \postcode{5095}
}
\email{mark.billinghurst@unisa.edu.au}

\renewcommand{\shortauthors}{Yang, et al.}

\begin{abstract}
We delineate the development of a mind-mapping system designed concurrently for both VR and desktop platforms. Employing an iterative methodology with groups of users, we systematically examined and improved various facets of our system, including interactions, communication mechanisms and gamification elements, to streamline the mind-mapping process while augmenting situational awareness and promoting active engagement among collaborators. We also report our observational findings on these facets from this iterative design process.
\end{abstract}

\ccsdesc[500]{Human-centered computing~User studies}
\ccsdesc[500]{Human-centered computing~Interaction devices}
\ccsdesc[500]{Human-centered computing~Collaborative and social computing systems and tools}
\ccsdesc[500]{Human-centered computing~Empirical studies in collaborative and social computing}

\keywords{Collaborative Sensemaking, Virtual Reality, Gamification, Hand Gestures}

\begin{teaserfigure}
  \centering
  \includegraphics[width=0.4325\textwidth]{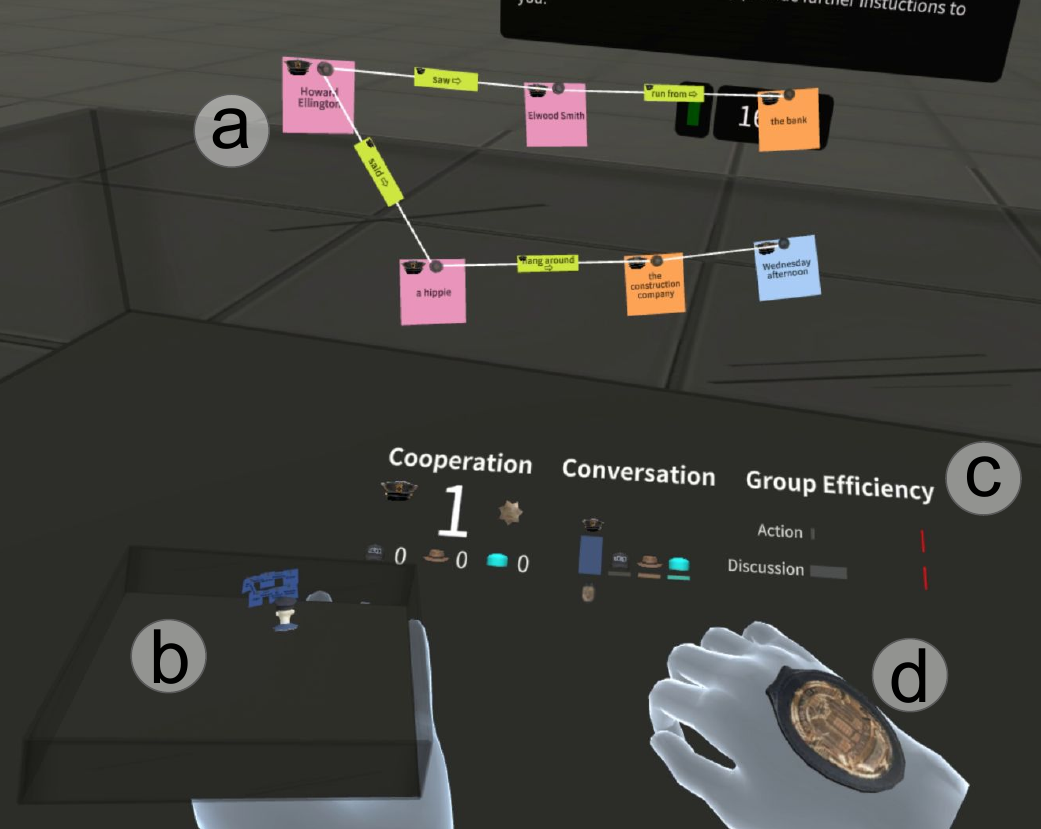}
  \includegraphics[width=0.5575\textwidth]{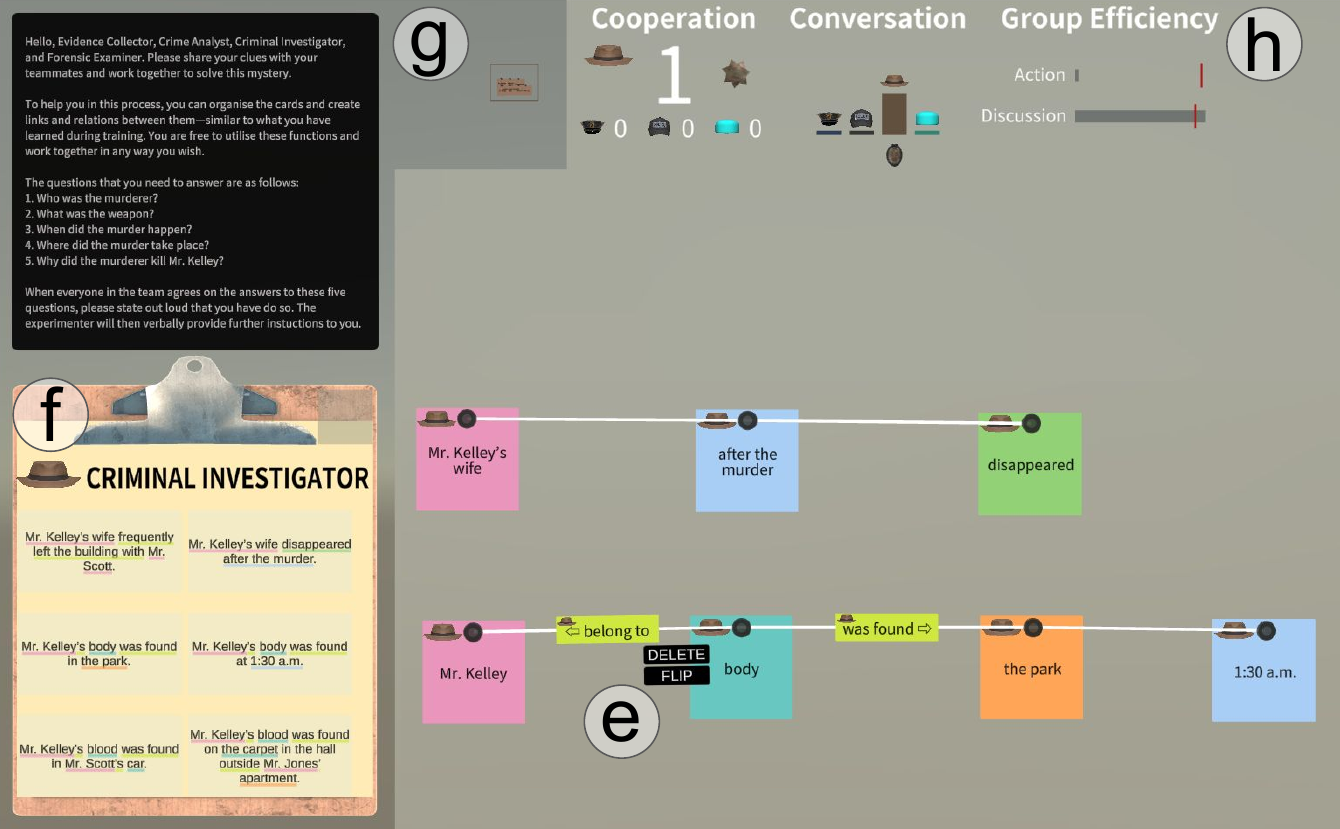}
  \caption{
  The final VR system (left) and desktop interface (right) that we developed after four iterations.
  (a) Notes with pins, and links with labels attached to the notes. (b) A 3D minimap is shown above the left palm in VR (the clipboard on the inner side of the palm is hidden to save room), with avatar, notes and links coloured the same as the role colour. (c) The gamification board in VR is attached to the body at waist height. (d) The embodied badge is on the back of the hand. (e) Buttons are shown for a note while selected. (f) Clipboard on the desktop interface. (g) Minimap on the desktop interface, with a coloured square showing the current field of view. (h) Gamification board on the desktop interface.
  }
  \label{fig:teaser}
\end{teaserfigure}


\maketitle

\section{Introduction}

Mind maps are visual externalisations of thoughts that can be used to extend working memory and synthesise ideas in collaborative sensemaking tasks~\cite{buzan1983use,mahyar2014supporting}. 
Desktop video conferencing and collaborative digital tools such as Miro\footnote{\url{https://miro.com/}} are now commonplace tools to support mind mapping, however, emerging immersive technologies such as virtual reality (VR) foster whole-body presence and provide natural interaction, resembling in-person experiences more closely than desktop interfaces~\cite{witmer1998measuring,raja2004exploring,yang2022towards}.

Limited investigation has been conducted regarding mind-mapping tools in immersive environments~\cite{miyasugi2017implementation,kutak2019interactive,lee2021post,tong2023towards}.
Adopting the Rapid Iterative Test and Evaluation (RITE) methodology~\cite{medlock2005rapid}, we contribute an iterative design and development of a VR and desktop mind-mapping system tailored to facilitate collaborative sensemaking and problem-solving, with innovative features to enhance situational awareness, spatial placement and active engagement, alongside intriguing experimental observations and findings.

\section{System Design Iterations}

In the four iterations conducted, we tested the baseline system in the 1st iteration, enhanced situation awareness in the 2nd, improved spatial placement in the 3rd and experimented with gamification in the 4th iteration. We recruited two groups of four users in each iteration.

\subsection{Baseline System Design}

Mind maps normally consist of labelled nodes connected by links. 
Our system uses virtual sticky notes to represent nodes, metaphoric to physical sticky notes. These notes, available in various colours, can be freely positioned in the 3D environment in VR (see Figure~\ref{fig:teaser} left) and placed on a 2D canvas in the desktop system like on a Miro board (see Figure~\ref{fig:teaser} right). 
Each note features a pin to which links are attached (see Figure~\ref{fig:teaser}a). Link labels are represented by narrow rectangular markers containing textual arrows indicating their direction. 
Multiple labels can be affixed to a single link, aligned vertically along its midpoint.

\begin{figure*}
    \centering
    \includegraphics[width=\textwidth]{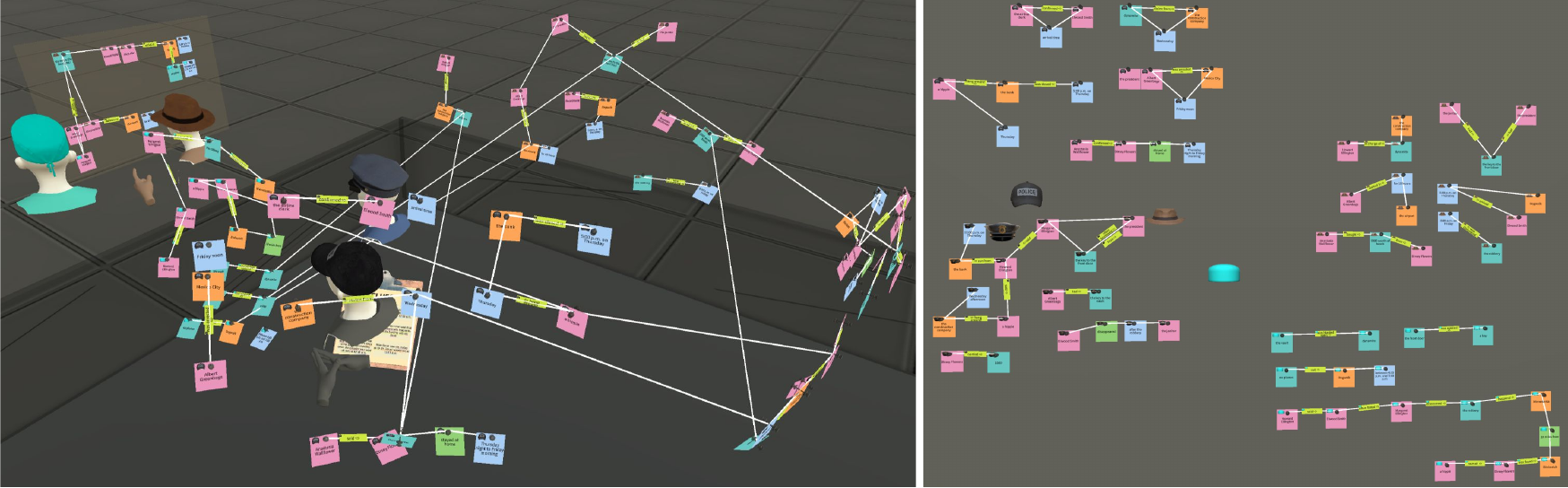}
    \caption{Examples of users creating mind maps in the shared workspace of the VR (left) and desktop (right) system.}
    \label{fig:study}
\end{figure*}

The space functions as a shared environment akin to a co-located workplace, facilitating synchronised updates to the mind map across all users.
In VR, users are represented by 3D avatars displaying head, upper body, and hands (see Figure~\ref{fig:study} left), accompanied by name tags positioned above their heads and oriented towards viewers. 
Avatar movements and hand gestures align with users' real-world actions and are synchronised for all users. Voice transmission is accomplished with lip sync applied to avatar, and 3D spatial sound adjusting volume based on proximity.
In the desktop system, users are depicted as 2D avatars, labelled with their names, 
indicating mouse cursor positions (see Figure~\ref{fig:study} right). Avatar movement and voice transmission are consistent with the VR setup, albeit with non-spatial audio effects.

To ensure uninterrupted workflow, we implemented intuitive physical movements for navigation, avoiding virtual locomotion methods that may disrupt cognitive processes.
In the desktop application, users navigate the canvas with panning by left-clicking and dragging the mouse, and zooming with the scroll wheel.

\subsection{Interaction Design}

\begin{figure*}
    \centering
    \includegraphics[width=\textwidth]{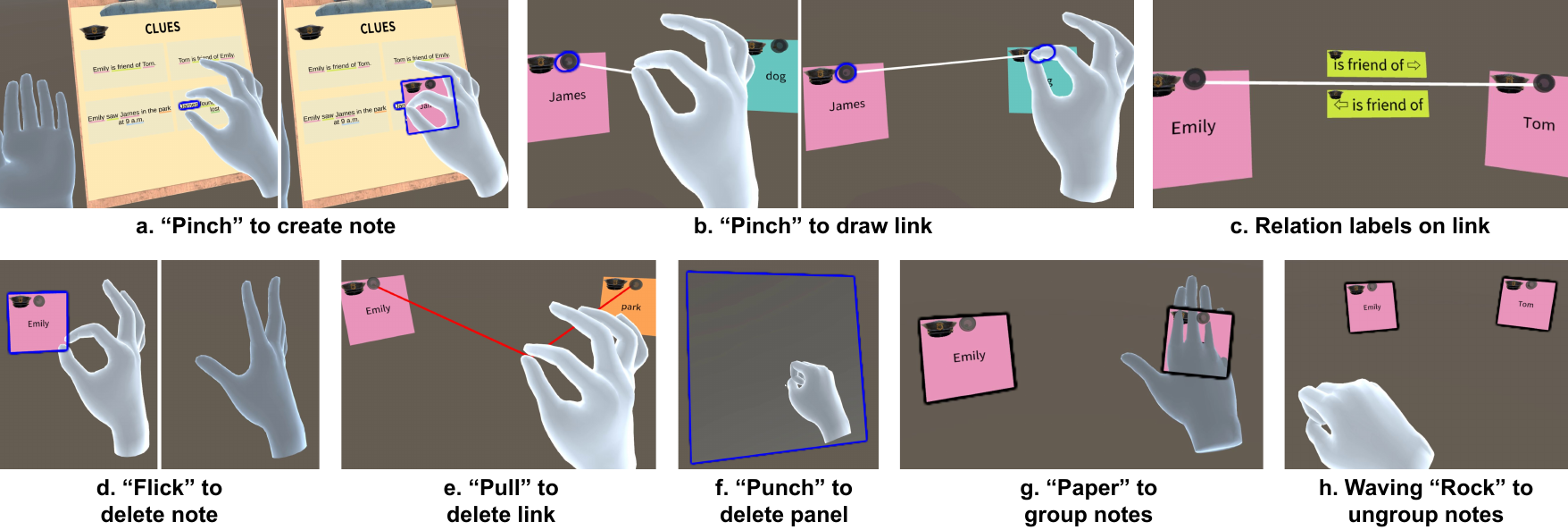}
    \caption{Hand gestures in the VR system.}
    \label{fig:gestures}
\end{figure*}

To expedite mind map editing in VR, we rely solely on hand gestures, eschewing controllers and traditional menus. These gestures, informed by previous research~\cite{piumsomboon2013user,huang2017gesture}, are designed to be intuitive and natural. 
Notably, we employ a ``pinch'' gesture with the index and middle fingers for grabbing virtual objects, including creating sticky notes and labels by dragging text from a clipboard, which is shown when the hand is flattened with the palm up (see Figure~\ref{fig:gestures}a). 
Linking notes involves pinching and dragging from one note's pin to another (see Figure~\ref{fig:gestures}b). 
Labels are attached to links by dropping label onto the desired link, with detaching them by dragging away. 
Flipping a label can alter a link's direction (see Figure~\ref{fig:gestures}c).
Deleting notes is achieved through a ``flicking'' gesture (see Figure~\ref{fig:gestures}d) and deletion of links is performed with ``pulling'' (see Figure~\ref{fig:gestures}e). These deleting gestures have been refined over iterations, experimenting with ``scissor'', ``crumpling'' and ``slicing'', discussed in Section~\ref{sec:findings}.
Multi-selection of notes are facilitated by a ``grouping'' gesture with a flattened hand (see Figure~\ref{fig:gestures}f). 
Waving a ``rock'' gesture in the air ungroups notes (see Figure~\ref{fig:gestures}g). 

In the desktop system, interactions are mouse-driven, including dragging objects from the clipboard and creating links by clicking and dragging pins. Deletion and flipping of labels are accomplished via dedicated buttons (see Figure~\ref{fig:teaser}e). Shift-clicking groups/ungroups notes.

In both platforms, visual cues and sound effects accompany interactions for enhanced feedback to all collaborators.

\subsection{Situational Awareness}

To better support situational awareness during collaboration to notify others' works, we added ``clone indicator'' to help users identify the same nodes across mind maps built by collaborators.
A clone indicator is shown when a user creates/grabs a sticky note, the content of which already exists in the workspace. The clone indicators are red tapered links~\cite{holten2011extended}, originate from the centre of the grabbed note and extend to all other identical notes. 
The colour intensity of the links is animated to fade in and out to attract attention.

To enhance the visibility of sticky note creators and provide an overview of individual workspaces, we devised a miniature of the working area. 
This miniature depicts virtual objects such as notes, links and labels, as well as users' avatars. The virtual objects are colour-coded as their creator's avatar colour. 
In the VR system, the miniature is displayed atop the palm centre and appears when a user flattens their hand. As users rotate their bodies in the VR environment, the miniature adjusts its orientation on the palm to align with user's forward direction (see Figure~\ref{fig:teaser}b).

\subsection{Spatial Placement}

To support the placement and arrangement of sticky notes, we added 2D panels.
A 2D panel serves as a surface for attaching and detaching notes. 
In VR, the panel is designed as a transparent light grey "glass" to allow visibility from both sides and easy differentiation from the sky. 
Users can retrieve a panel from the clipboard via a designated icon in the upper-right corner. 
Panels can accommodate multiple notes and can be freely moved using a grabbing gesture. 
A "snap" zone facilitates the automatic attachment of notes to the panel when within close proximity. 
Panels dynamically adjust their size to accommodate attached notes, and this resizing also can be done manually. 
Panels can be deleted in VR through a punching gesture resembling smashing glass (see Figure~\ref{fig:gestures}f), while in the desktop system, deletion is achieved by clicking the "DELETE" button.

\subsection{Gamification}

To improve users' engagement and efficiency, we adopted gamification mechanism and designed certain measures and a visualisation dashboard to automatically give real-time feedback on users' performance.
To encourage users to work together in action, we quantified two cooperation behaviours: linking two sticky notes created by others, or attaching a label to a link created by others; to promote engagement in discussion, we tracked users' speaking duration. 
We also assessed efficiency in actions and discussions, which was defined as the ratio of produced output and spent input~\cite{del2011measuring}, showing how fast users finish the construction of mind maps, or how much time users spend on discussion during problem-solving.
These measures are visualised as bar charts on the dashboard, which are updated in real-time and attached to the avatar body at waist-high in front of users in VR (see Figure~\ref{fig:teaser}c) or on the top of the screen on desktop, with game sound effects signalling score changes or leadership shifts. A badge is also given to the leading users (see Figure~\ref{fig:teaser}d).

\section{Observations \& Discussions}
\label{sec:findings}

From our observation, ``clone indicator'' prompted users to build one mind map and work together. Users instinctively started building mind maps after this feature was introduced. 

Regarding the miniature, users commented that it was fun to look at especially in VR, but not useful as the space was small.
While \cite{yang2020embodied} found that an overview provided benefits over physical movement in 3D scatterplots analysis in a $2\times2$ meters space in VR, we suspect that the constant need to modify and manipulate mind maps in our scenario made the noninteractable overview minimap less useful than it in more passive activities and bigger space.

We were aware of the benefits of 2D panels for spatial arrangement from previous work~\cite{lee2021post,tahmid2022evaluating}. In our observation, interestingly 2D panels were used in various ways: (1) to categorise information, (2) to ``store'' important information, (3) to visually ``merge'' duplicated notes, and (4) to claim working territories.

We observed positive effects of gamification. 
In summary, our results showed that users were stimulated by the designed gamification to construct mind maps together and find an efficient working strategy.
Also, while users did not discuss the gamification scores explicitly during tasks, they showed a clear difference in strategy and effort-demanding tasks than those without gamification.

From our iterative gesture refinement in VR, we draw insights beyond  prior research~\cite{piumsomboon2013user,miyasugi2017implementation,huang2017gesture,newbury2021embodied,olaosebikan2022embodied}:
(1) Natural finger movements yield superior results compared to unnatural counterparts. 
The "scissor" gesture garnered negative feedback due to reported fatigue and discomfort, whereas gestures like "crumpling" and "flicking" received no such complaints, as their finger movements felt more natural;
(2) Gestures not reliant on depth perception outperform those that do. 
Users struggled with the "slicing" gesture for object deletion, often failing to realise they made the gesture without touching the object, despite our provision of a blue outline for touch notification.

\section{Conclusion}

We introduce a collaborative mind-mapping system accessible in both VR and 2D desktop environments. Utilising the RITE methodology, we conducted four iterations to develop and refine our system to enhance natural interactions, situational awareness, spatial placement and active engagement in collaboration.


\bibliographystyle{ACM-Reference-Format}
\bibliography{main}


\end{document}